\documentclass[12pt]{iopart}

\usepackage{iopams}
\usepackage{graphicx,graphics,color,epsfig}

\newcommand{\ket}[1]{| #1 \rangle}

\begin{document}

\title[CEWQO 18]{Entanglement transfer between bipartite systems}

\author{Smail Bougouffa$^{1}$ and Zbigniew Ficek$^{2}$}
\address{$^{1}$Department of Physics, Faculty of Science, Taibah University, P.O. Box 30002, Madinah, Saudi
Arabia\\ $^{2}$The National Center for Mathematics and Physics,
KACST, P.O. Box 6086, Riyadh~11442, Saudi Arabia} \ead{
sbougouffa@hotmail.com, sbougouffa@taibahu.edu.sa}

\begin{abstract}
The problem of a controlled transfer of an entanglement initially encoded into two two-level atoms that are successively sent through two single-mode cavities is investigated. The atoms and the cavity modes form a four qubit system and we demonstrate under which conditions the initial entanglement encoded into the atoms can be completely transferred to other pairs of qubits. We find that in the case of a nonzero detuning between the atomic transition frequencies and the cavity mode frequencies, no complete transfer of the initial entanglement is possible to any of the other pairs of qubits. In the case of exact resonance and equal coupling strengths of the atoms to the cavity modes, an initial maximally entangled state of the atoms can be completely transferred to the cavity modes. The complete transfer of the entanglement is restricted to the cavity modes only with the transfer to the other pairs being limited to up to $50$\%. We have found that the complete transfer of an initial entanglement to other pairs of qubits may take place if the initial state is not the maximally entangled state and the atoms couple to the cavity modes with unequal strengths. Depending on the ratio between the coupling strengths, the optimal entanglement can be created between the atoms and one of the cavity modes.
\end{abstract}

\pacs{03.65, 03.67, 42.50}
\submitto{\PS}
\maketitle

\section{Introduction}

There has recently been a great interest in the study of entanglement, its generation~\cite{IBV99, GSG05, CP06, LMGQX07, JDS08}, evolution and storage~\cite{SFEDNH05,S07, CRRJ08, QJHC10}. An another interesting problem, crucial for the quantum information processing is a controlled transfer of an initial entanglement to a desired pair of qubits. The dynamics and a controlled transfer of entanglement between atoms located inside two completely isolated cavities have recently been investigated by Sainz and Bj\"ork~\cite{IG07},~Yonac and Eberly~\cite{ MJ06, MJ07, MJ10} and Chan \textit{et al.}~\cite{SMZ09,SMZ10}. It has been shown that a periodic and the direct transfer of an entanglement from a qubit to an another qubit could be achieved by a suitable engineering of the coupling strengths of the atoms to the cavity field. Some other interesting aspects of entanglement of atoms interacting with cavities have also been discussed. Particularly interesting are the studies of the evolution of entanglement under the influence of different reservoirs~\cite{S10,TMSZ10, SA11}.

In this paper, we extend these ideas to the case of atoms successively passing two single-mode cavities. The atoms and the cavity modes form a four qubit system and we examine the evolution of the possible six pairwise concurrences that can be distinguished in the system. We are particularly interested under which conditions the initial entanglement encoded into the atoms can be completely transferred to the other pairs of qubits. Simple analytical formulas are obtained for the pairwise concurrences and the dependence of the time evolution of the concurrences on the initial atomic conditions, detuning of the atomic transition frequencies from the cavity field frequencies and the ratio of the coupling strengths of the atoms to the cavity modes are analysed. We determine the parameter ranges in which the initial entanglement can be completely transferred to a particular pair of qubits. We find an interesting result that non-maximally entangled states are more robust for the complete transfer than maximally entangled states.

\section{Description of the system and Hamiltonian}

\begin{figure}[h]
\center{ \includegraphics[width=0.4\textheight]{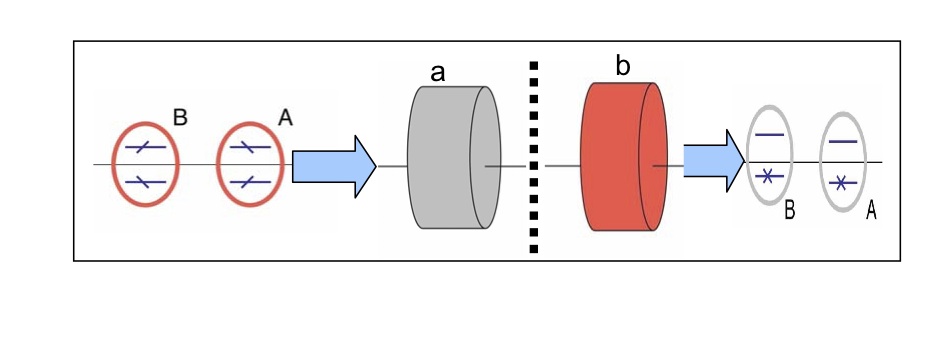}}
\caption{(Color online) Schematic diagram of the system composed of two single-mode cavities, $a$ and $b$, and two two-level atoms, $A$ and $B$, passing through the cavities. The atoms are initially prepared in an entangled single-excitation state while the cavities are in the vacuum state.} \label{bfig1}
\end{figure}

We consider a four qubit system composed of two single-mode cavities, labeled as~$a$ and~$b$, and two two-level atoms, labeled as $A$ and $B$. Each atom has energy levels~$\ket{g_{i}}$ and $\ket{e_{i}}\, (i=A,B)$ separated by frequency $\omega_{0}$ and connected by a transition dipole moment $\mu_{i}$. The cavities are modelled as composed of single bosonic modes of equal frequencies $\omega_{a}=\omega_{b}\equiv\omega_{c}$ and represented by operators $\hat{a}\, (\hat{b})$ and  $\hat{a}^{\dag}\, (\hat{b}^{\dag})$ which are, respectively, the annihilation and creation operators for the cavity $a$ $(b)$ mode. The atoms are initially prepared in a single-excitation entangled state and then are successfully directed through the cavities, as illustrated in figure~\ref{bfig1}. The cavity modes are initially in the zero temperature vacuum state. The atoms, when passing through the cavities, interact with the cavity modes with electric-dipole coupling strengths $g_{a}$ and $g_{b}$, respectively. The interaction of the atoms with the cavity fields leads to a transfer of the initial entanglement to one or few of the possible six pairwise subsystems,~$(AB), (Aa), (ab), (Ba), (Ab)$ and $(Bb)$. To which subsystem the initial entanglement has been transferred is quantified by calculating concurrences~$C_{AB}, C_{Aa}, C_{ab}, C_{Ba}, C_{Ab}$ and~$C_{Bb}$.

The total Hamiltonian of the system in the electric-dipole and rotating-wave approximations can be written as
\begin{eqnarray}
H_{T} &=& \frac{1}{2}\hbar\omega_{0}\left(\sigma^{z}_{A}+\sigma^{z}_{B}\right) +\hbar\omega_{c}\hat{a}^{\dag}\hat{a} +\hbar\omega_{c}\hat{b}^{\dag}\hat{b} \nonumber\\
&+& \hbar g_{a}\left(\sigma_{A}^{+}\hat{a} +\sigma_{B}^{+}\hat{a} +{\rm H.c.}\right) +\hbar g_{b}\left(\sigma_{A}^{+}\hat{b} +\sigma_{B}^{+}\hat{b} +{\rm H.c.}\right) ,\label{1}
\end{eqnarray}
where $\sigma_{i}^{+}$ and $\sigma_{i}^{-}$ are, respectively, the raising and lowering operators of atom $i$, and~$\sigma^{z}_{i}$ describes its energy. In writing the Hamiltonian (\ref{1}), we have assumed that the cavities are close enough to each other that the retardation effects in the interaction of the atoms with the cavity modes can be ignored. We shall further assume that the travelling time of the atoms through the cavities is short compared to the lifetime of the atoms and the cavity excitation so that spontaneous decay of the atoms and damping of the cavity modes can be neglected.

We now proceed to examine the process of transferring an initial entanglement encoded into the atoms to a desired pair of qubits. Before the first atom enters the cavities, the space of the system is spanned by four product state vectors defined~as
\begin{eqnarray}
\ket 1 &=& \ket{e_{A}}\otimes\ket{g_{B}}\otimes\ket{0_{a}}\otimes\ket{0_{b}} ,\quad \ket 2 =\ket{g_{A}}\otimes\ket{e_{B}}\otimes\ket{0_{a}}\otimes\ket{0_{b}} , \nonumber\\
\ket 3 &=& \ket{g_{A}}\otimes\ket{g_{B}}\otimes\ket{1_{a}}\otimes\ket{0_{b}} ,\quad \ket 4 =\ket{g_{A}}\otimes\ket{g_{B}}\otimes\ket{0_{a}}\otimes\ket{1_{b}} .\label{2}
\end{eqnarray}
where $\ket{e_{A}}\otimes\ket{g_{B}}\otimes\ket{m_{a}}\otimes\ket{n_{b}}$ stands for the atom $A$ in the excited state, the atom $B$ in the ground state, $m$ photons in the cavity $a$, and $n$ photons in the cavity $b$, respectively.

When the atoms are passing through the cavities, the state vector of the system~$\ket{\Psi(t)}$ evolves in time, and its evolution is governed by the Schr\"odinger equation
\begin{eqnarray}
  i\hbar\frac{d}{d t}\ket{\Psi(t)} = \tilde{H}\ket{\Psi(t)} ,\label{3}
\end{eqnarray}
where $\tilde{H}$ is of the Hamiltonian of the system in the interaction picture. Since $\tilde{H}$ does not depend explicitly on time, the Schr\"odinger equation (\ref{3}) has the formal solution
\begin{eqnarray}
  \ket{\Psi(t)} &=& \exp\left[-i(\tilde{H}/\hbar)t\right]|\Psi(0)\rangle \nonumber\\
  &=& \alpha_{1}(t)\ket 1 + \alpha_{2}(t)\ket 2 + \alpha_{3}(t)\ket 3 + \alpha_{4}(t)\ket 4 ,\label{4}
\end{eqnarray}
where the time depended probability amplitudes are
\begin{eqnarray}
  \alpha_{1}(t) &=& u_{0} +w_{0}\,{\rm e}^{i\Delta t}\left[\cos(\Omega t)-i\frac{\Delta}{\Omega}\sin(\Omega t)\right] ,\nonumber\\
  \alpha_{2}(t) &=& -u_{0} +w_{0}\,{\rm e}^{i\Delta t}\left[\cos(\Omega t)-i\frac{\Delta}{\Omega} \sin(\Omega t)\right] ,\nonumber \\
  \alpha_{3}(t) &=& -2i\frac{w_{0}g_{a}}{\Omega} {\rm e}^{i\Delta t}\sin(\Omega t) ,\nonumber \\
  \alpha_{4}(t) &=& -2i\frac{w_{0}g_{b}}{\Omega} {\rm e}^{i\Delta t}\sin(\Omega t) ,\label{5}
\end{eqnarray}
in which $u_{0}$ and $w_{0}$ describe initial conditions at $t=0$, the parameter $2\Delta =\omega_0-\omega_{c}$ is the detuning of the atomic transition frequency from the cavity field frequency, and $\Omega= \sqrt{\Delta^2+2(g_{a}^{2}+g_{b}^{2})}$ is a detuned Rabi frequency. For the initial state of the system, we choose a product state $\ket{\Psi(0)} = (\cos\theta \ket{e_{A},g_{B}} +\sin\theta \ket{g_{A},e_{B}})\otimes\ket{0_{a}}\otimes\ket{0_{b}}$ with an entangled state for the atoms, determined by the angle $\theta$, and zero occupation numbers for the cavity modes. In this case, $u_{0}=(\cos\theta -\sin\theta)/2$ and~$w_{0}=(\cos\theta +\sin\theta)/2$.

\section{Entanglement transfer}

Given the state of the system, it is straightforward to calculate concurrences~\cite{W98} of the distinguished pairs of qubits and to analyse conditions for the complete transfer of the initial entanglement to a desired pair of qubits. It is easy to show that the concurrences can be written in terms of the probability amplitudes as
\begin{eqnarray}
 C_{AB}(t) &=& 2\left|\alpha_{1}(t)\alpha_{2}^{\ast}(t)\right| ,\quad C_{ab}(t) =2\left|\alpha_{3}(t)\alpha_{4}^{\ast}(t)\right| , \nonumber\\
 C_{Aa}(t) &=& 2\left|\alpha_{3}(t)\alpha_{1}^{\ast}(t)\right| ,\quad  C_{Ab}(t) =2\left|\alpha_{4}(t)\alpha_{1}^{\ast}(t)\right| ,\nonumber\\
 C_{Ba}(t) &=& 2\left|\alpha_{2}(t)\alpha_{3}^{\ast}(t)\right| ,\quad C_{Bb}(t) =2\left|\alpha_{2}(t)\alpha_{4}^{\ast}(t)\right| .\label{6}
\end{eqnarray}
The time evolution of the concurrences depends on three parameters, the angle $\theta$ that determines the degree of entanglement of the initial state, the detuning~$\Delta$ and the ratio of the coupling constants $g_{a}$ and $g_{b}$. For simplicity, we consider the case of exact resonance between the atoms and the cavity modes $(\Delta =0)$. In fact, it can be seen from (\ref{5}) that for nonzero detuning $(\Delta\neq 0)$ the concurrences oscillate between zero and a certain value that is less than one for all times $t>0$. Thus, in the case of $\Delta\neq 0$, an entangled initial state of the atoms cannot be completely transferred to the other pairs of the qubits. The entanglement redistributes over the qubits.

\begin{figure}[h]
\center{\includegraphics[width=0.45\textheight]{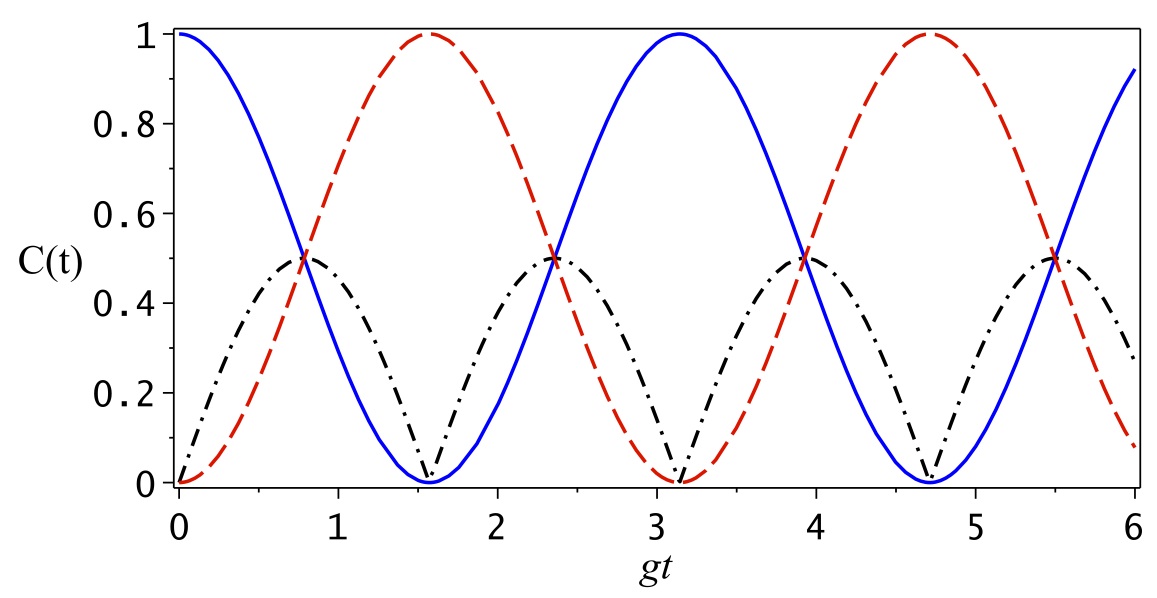}}
\caption{(Color online) The time evolution of the concurrences for the initial maximally entangled state between the atoms $C_{AB}(0)=1$, corresponding to $\theta=\pi/4$, and for equal coupling strengths $g_{a}=g_{b}=g$. The blue solid line is $C_{AB}(t)$, red dashed line is $C_{ab}(t)$, and black dashed-dotted line is $C_{Aa}(t)=C_{Ab}(t)=C_{Ba}(t)=C_{Bb}(t)$. Also shown is the sum of the squared concurrences $C_{AB}^{2}(t)+C_{Ab}^{2}(t)+C_{Ba}^{2}(t)+C_{ab}^{2}(t)$ (red solid line).}
\label{fig2}
\end{figure}

Let us first illustrate a special case of the transfer of an initial maximally entangled state of the atoms, corresponding to~$u_{0}=0, w_{0}=1/\sqrt{2}$, through the system with both atoms coupled equally to the cavity modes~$g_{a}=g_{b}=g$. One can see from (\ref{5}) that in this case $\alpha_{1}(t)=\alpha_{2}(t)$ and~$\alpha_{3}(t)=\alpha_{4}(t)$ for all times, so that the concurrences (\ref{6})~are
\begin{eqnarray}
 C_{AB}(t) &=& \cos^{2}(2gt) , \quad C_{ab}(t) = \sin^{2}(2gt) , \nonumber\\
  C_{Aa}(t) &=& C_{Ab}(t) = C_{Ba}(t) = C_{Bb}(t) =\frac{1}{2}\left|\sin (4gt)\right| .\label{7}
\end{eqnarray}
As can be seen from (\ref{7}), the maximal entanglement initially encoded into the atoms~$C_{AB}(0)=1$ can be completely transferred to the cavity modes, $C_{ab}(t_{n})=1$ at certain times satisfying~$t_{n}= n(\pi/4g),\, n=1,3,5,\ldots$. No complete transfer of the entanglement is possible to other pairs of qubits which, according to (\ref{7}), can be maximally entangled only up to $50$\%. This is illustrated in figure~\ref{fig2}, where we plot the time evolution of the concurrences for the initially maximally entangled atoms. The complete transfer of the initial entanglement is seen to occur only between the pairs~$C_{AB}(t)$ and $C_{ab}(t)$. The other pairs are identical and can be entangled to only $50$\% at times when the entanglement is evenly redistributed over all pairs of qubits.

Figure~\ref{fig2} also shows the sum of the squares of the concurrences that $C_{AB}^{2}(t)+C_{Ab}^{2}(t)+C_{Ba}^{2}(t)+C_{ab}^{2}(t)=1$ for all times. This result exemplifies some of the remarks made by Yonac \textit{et al.}~\cite{MJ07} and Chen \textit{et al.}~\cite{SMZ09} in connection with the conservation of the total square of the concurrence. Namely, similar to the case of separate cavities each containing a single atom, the sum of the squared concurrences in the system of two initially entangled atoms passing through both cavities is conserved.

\begin{figure}[h]
\center{\includegraphics[width=0.45\textheight]{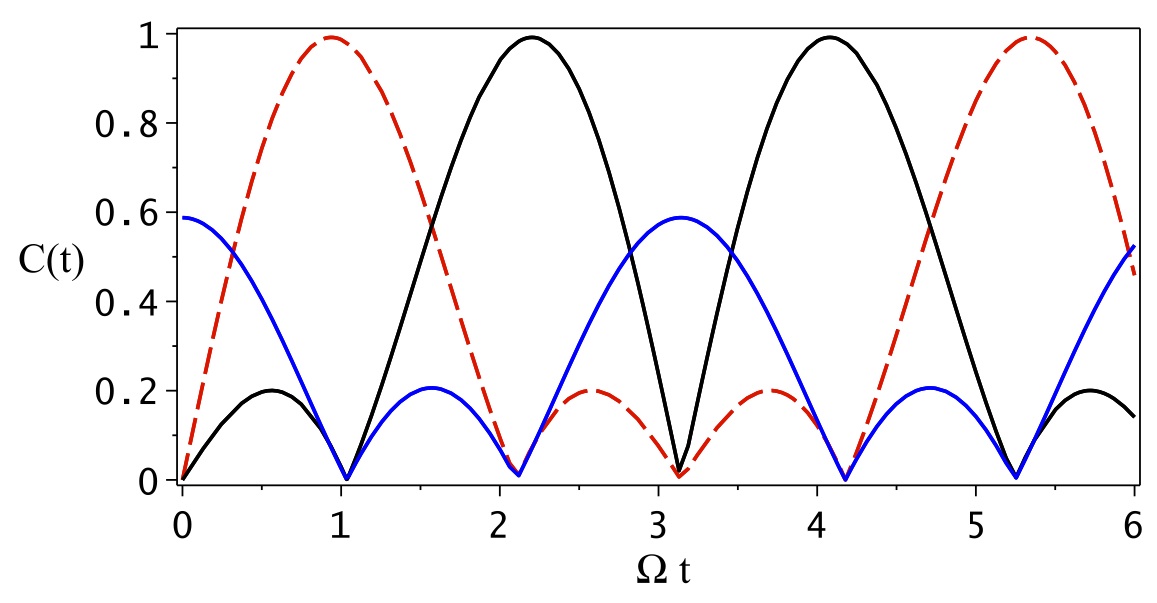}}
\caption{(Color online) The time dependence of the concurrences for an initial non-maximally entangled state of the atoms with $\theta=\pi/10$, $\Delta=0$ and $g_b/g_a=0.1$. The blue solid line shows $C_{AB}(t)$, the red dashed line shows $C_{Aa}(t)$, and the black solid line shows $C_{Ba}(t)$.}
\label{fig3}
\end{figure}

We now present some numerical calculations that illustrate the transfer process of an initially non-maximally entangled state of the atoms to other pairs of the qubits. It turns out that in this case, the complete transfer of an initial entanglement could occur only for unequal coupling strengths $g_{a}\neq g_{b}$. This interesting behaviour of the entanglement evolution is illustrated in figure~\ref{fig3}, where we plot the time evolution of the concurrences  $C_{AB}(t), C_{Aa}(t)$ and $C_{Ba}(t)$ for the initial non-maximally entangled state of the atoms with $\theta=\pi/10$ and $g_b/g_a=0.1$. The remaining concurrences $C_{Ab}(t), C_{Bb}(t)$ and $C_{ab}(t)$ are found to be close to zero for all times and are not displayed in the figure. It is seen from the figure that the concurrences oscillate in time with periodically modulated amplitudes and at certain discrete times, the concurrences $C_{Aa}(t)$ and $C_{Ba}(t)$ reach the maximal values close to unity. The maxima of $C_{Ba}(t)$ are however displaced relative to those of $C_{Aa}(t)$.

To examine the features of $C_{Aa}(t)$ and $C_{Ba}(t)$, seen in figure~\ref{fig3}, we analyse of the expressions (\ref{6}) for the concurrences which with the probability amplitudes~(\ref{5}) show that at zero detuning and for a strongly asymmetric coupling $g_a\gg g_b$, the concurrences~$C_{Ab}(t), C_{Bb}(t)$ and~$C_{ab}(t)$ are close to zero for all times, whereas the concurrences~$C_{Aa}(t)$ and $C_{Ba}(t)$ can be written as
\begin{eqnarray}
 C_{Aa}(t) &=& \frac{4g_a w_{0}^2}{\Omega}\left|\sin(\Omega t)\!\left[\frac{(u_{0}-w_{0})}{w_{0}} +2\cos^2\left(\frac{1}{2}\Omega t\right)\right]\right| ,\nonumber\\
 C_{Ba}(t) &=& \frac{4g_a w_{0}^2}{\Omega}\left|\sin(\Omega t)\!\left[\frac{(u_{0}-w_{0})}{w_{0}} +2\sin^2\left(\frac{1}{2}\Omega t\right)\right]\right| .\label{8}
\end{eqnarray}
It is readily found from (\ref{8}) that the concurrence~$C_{Aa}(t)$ is greatest at times $t_{n}=n(\pi/3\Omega),\, n=1,5,7,11,\ldots$, and reaches the maximum value that is close to unity when $|u_{0}-w_{0}|\ll w_{0}$. On the other hand, the time variation of $C_{Ba}(t)$ is phase shifted relative to $C_{Aa}(t)$. Therefore, the positions of the maxima of $C_{Ba}(t)$ are displaced relative to those of $C_{Aa}(t)$ and appear at times $t_{m}= m(\pi/3\Omega),\, m=2,4,8,10,\ldots$, the kind of behaviour seen in figure~\ref{fig3}. 

We would like to point out that the effect of going to the transfer of non-maximally entangled states through unequal couplings of the atoms to the cavity modes is clearly to achieve not only the complete transfer of the initial entanglement but also to create the maximal concurrences~$C_{Aa}(t)$ and $C_{Ba}(t)$. The reason is that during the passing of a given atom, either $A$ or~$B$ through the cavity $a$, the interaction of the atom with the field leads not only to the transfer of the initial entanglement from the atom to the cavity field but also creates an additional entanglement. Even in the limit of no initial entanglement in the system, nonzero concurrences $C_{Aa}(t)$ and $C_{Ba}(t)$ can be generated during the passage of the atoms through the cavity $a$. A similar conclusion applies to the case of $g_b\gg g_a$, where the maximal concurrences $C_{Ab}(t)$ and $C_{Bb}(t)$ can be created between the atoms and the field mode of the cavity $b$.

\section{Conclusions}

We have studied the problem of the entanglement transfer between different pairs of qubits in a system composed of four qubits, two two-level atoms plus two single-mode cavities. We have shown that in the case of a nonzero detuning between the atomic transition frequencies and the cavity mode frequencies, no complete transfer of an initial entanglement is possible to any of the six possible pairs of qubits. In the case of zero detuning and equal coupling strengths of the atoms to the cavity modes, an initial maximally entangled state of the atoms can be completely transferred to a pair of qubits involving the cavity modes. The complete transfer of the entanglement is restricted to the cavity modes with the transfer to the other pairs being limited to up to $50$\% only. We have found that the complete transfer of an initial entanglement to other pairs of qubits may take place if the initial state is not the  maximally entangled state and the atoms couple to the cavity modes with unequal strengths. In this case the optimal entanglement can also be created between some of the pairs of qubits.

\end{document}